\documentclass[a4paper,prd,11pt,noshowkeys,notitlepage,nofootinbib,superscriptaddress]{revtex4}
\pdfoutput=1  %necessary for JHEP
\usepackage[utf8]{inputenc}
\usepackage{ae,aecompl}
\usepackage{amsmath,amssymb,mathrsfs}
\usepackage{graphicx}
\usepackage{ccnishi}
\usepackage{dsfont}
\usepackage{slashed}
\usepackage{units}
\usepackage[normalem]{ulem}

\usepackage[usenames,dvipsnames]{xcolor} % svgnames
\usepackage{hyperref}
\hypersetup{
    colorlinks=true,        % false: boxed links; true: colored links
    linkcolor=purple,       % color of internal links, e.g., eqs.
    citecolor=JungleGreen,  % color of links to bibliography
    filecolor=magenta,      % color of file links
}

\providecommand{\cM}{\mathscr{M}}

\providecommand{\cY}{\mathscr{Y}}
\providecommand{\cO}{\mathcal{O}}
\providecommand{\cH}{\mathcal{H}}

\providecommand{\tw}{\tilde{w}}

\providecommand{\tH}{\tilde{H}}

\providecommand{\btheta}{\bar{\theta}}

\providecommand{\bz}{\mathbf{z}}
\providecommand{\br}{\mathbf{r}}

\providecommand{\hY}{\hat{Y}}
\providecommand{\hX}{\hat{X}}
\providecommand{\hcY}{\hat{\cY}}

\providecommand{\lag}{\mathscr{L}}

\providecommand{\eps}{\epsilon}

\makeatletter
\newcommand*\bigcdot{\mathpalette\bigcdot@{.5}}
\newcommand*\bigcdot@[2]{\mathbin{\vcenter{\hbox{\scalebox{#2}{$\m@th#1\bullet$}}}}}
\makeatother

\begin{document}
%%%%%%%%%%%%%%%%%%%%%%%%%%%%%%%%%%%%%%%%%%%%%%%%%
\title{
{\huge{
Nelson–Barr Models with \\
Vector-Like Quark Doublets
}}
}
\author{G.~H.~S.~Alves}
 \thanks{ alves.gustavo@ufabc.edu.br}
 \affiliation{Centro de Ci\^{e}ncias Naturais e Humanas,\\
 Universidade Federal do ABC, 09.210-170,
 Santo Andr\'{e}-SP, Brasil}
\author{C.~C.~Nishi}
 \thanks{E-mail: celso.nishi@ufabc.edu.br}
 \affiliation{Centro de Matem\'{a}tica, Computa\c{c}\~{a}o e Cogni\c{c}\~{a}o,
 Universidade Federal do ABC, 09.210-170,
 Santo Andr\'{e}-SP, Brasil}
\author{L.~Vecchi}
 \thanks{E-mail: luca.vecchi@pd.infn.it}
 \affiliation{INFN, Sezione di Padova,
Via Marzolo 8, Padova, 35131, Italy}

% \date{\today}
%%%%%%%%%%%%%%%%%%%%%%%%%%%%%%%%%%%%%%%%%%%%%%%%%
\begin{abstract}

We investigate Nelson--Barr solutions to the strong CP problem in which spontaneous CP violation is transmitted to the 
Standard Model through mixing with a vector-like partner of the SM quark doublet. We show that these constructions 
constitute compelling and phenomenologically viable alternatives to the more widely studied singlet-based NB models. A 
key result of our analysis is that an accidental symmetry of the renormalizable theory delays the leading contributions 
to $\bar\theta$ until three loops, naturally suppressing hadronic CP violation. 
We outline the main phenomenological constraints, including
future EDM experiments, as well as the main differences between these 
scenarios and generic models with doublet vector-like quarks.

\end{abstract}
%%%%%%%%%%%%%%%%%%%%%%%%%%%%%%%%%%%%%%%%%%%%%%%%%
% \pacs{12.60.Fr, 14.80.Cp, 11.30.Qc, 02.20.-a}
%\keywords{ }
%\twocolumn
\maketitle
%  \tableofcontents
%%%%%%%%%%%%%%%%%%%%%%%%%%%%%%%%%%%%%%%%%%%%%%%%%

\section{Introduction}
\label{sec:intro}

The persistent absence of observable CP violation in the strong interactions remains one of the most striking puzzles in 
particle physics. Among the proposed resolutions, the Nelson--Barr (NB) mechanism\,\cite{nelson,barr}
occupies a prominent position as an axion-less class of approaches. In this paradigm, CP is imposed as a 
fundamental symmetry of nature and is subsequently broken spontaneously by the condensate of a (fundamental or 
composite) scalar field. The resulting CP violation is transmitted to the Standard Model (SM) via fermionic mediators. A 
Yukawa interaction among the CP-violating order parameter, the SM fermions, and heavy vector-like mediators ensures 
that physical CP-violating phases enter the CKM matrix at tree level, while the QCD angle $\bar\theta$ is generated 
only radiatively. The radiative corrections involving the couplings of the CP-violating Higgs sector, which generically arise already at one-loop\,\cite{BBP,dine,effective.NB}\footnote{%
They can be postponed to two-loops using a nonconventional CP symmetry\,\cite{cp4}.
}, are model-dependent but can naturally lie well within current bounds, especially in models with a composite CP-violating 
scalar\,\cite{vecchi.2,Csaki:2025ikr}. There are, however, additional radiative contributions to $\bar\theta$ that arise solely from 
the minimal 
ingredients of these scenarios --- the SM plus the fermionic mediators \cite{Valenti:2021rdu}. These irreducible 
contributions remain under control provided an appropriate choice of field content is made. Because they are 
unavoidable, they offer a distinctive signature of such scenarios: NB models necessarily predict a nonzero $\bar\theta$, 
potentially within reach of future electric dipole moment (EDM) experiments.

In this work, we investigate a class of NB models in which CP violation is communicated via mixing between the SM quark 
doublets and vector-like quark doublets 
with the same electroweak charges
--- a scenario referred to as $q$-mediated NB models in 
Ref.~\cite{Valenti:2021rdu}.
This setup is a variant of the popular $d$-mediated NB 
framework (see 
examples in \cite{BBP,dine,Asadi:2022vys,consequences,nb-vlq:fit,nbvlq:more,d-mediation:recent,d-mediation:others}) where the mediators are vector-like quarks 
with the quantum numbers of the SM quark singlet $d$.
The case of $u$-mediation is analogously possible.
Two important differences distinguish these scenarios from those 
previously studied in the literature, motivating a dedicated analysis. First, generic realizations of this type of 
mediation are typically disfavored because they induce large two-loop contributions to $\bar\theta$~\cite{vecchi.14}. 
However, we show that minimal {\emph{renormalizable}} NB models with $q$-mediation are protected from these large corrections by 
an accidental symmetry. Second, it is well known that NB scenarios require special scrutiny because their coupling 
structure is so constrained that reproducing the CKM matrix and the observed Yukawa hierarchy becomes highly nontrivial 
and must be carefully verified. This feature marks a key difference from more common singlet vector-like quark 
(VLQ) models, which possess a less restrictive structure to fit the SM.

Our analysis is structured as follows. In Sec.~\ref{sec:NB} we review the basic elements of Nelson--Barr models with $q$-mediation and compare their flavor structure to that of generic VLQ scenarios. Details of the precise parameter mapping are presented in Appendix~\ref{ap:change}. In Sec.~\ref{sec:SM} we explore in detail the region of parameter space that reproduces the flavor structure and CP violation of the SM. This is a nontrivial task that can unfortunately only be carried out numerically. Our analysis is partially supported by analytical manipulations (collected in Appendices~\ref{ap:inversion} and~\ref{ap:max:mu}) and partially guided by basis-independent considerations (see Appendix~\ref{sec:invs}). Our scenarios are subject to several constraints, the most important of which are discussed in Sec.~\ref{sec:constraints}. These include perturbativity requirements (Sec.~\ref{sec:pert}), precision electroweak constraints (Sec.~\ref{sec:S.T}), and bounds from flavor-violating observables (Sec.~\ref{sec:FV}). Special attention is devoted to the non-observation of hadronic CP violation. In particular, the radiative corrections to the QCD $\bar\theta$ parameter are estimated in Sec.~\ref{sec:invariants}, as these do not decouple when the new physics scale is increased and thus provide a robust signature of NB scenarios with $q$-mediation. Our conclusions are presented in Sec.~\ref{sec:conclusions}.

%%%%%%%%%%%%%%%%%%%%%%%%%%%%%%%%%%%%%%%%%%%
\section{Doublet vector-like quark of Nelson-Barr type}
\label{sec:NB}

Let us introduce a Dirac quark $Q=Q_L+Q_R$ in the representation $(\bs{2},1/6)$ of the electroweak $SU(2)_L\times U(1)_Y$, the same of the SM quark doublets $q_{iL}$ (with $i=1,2,3$ the generation index).

A doublet vector-like quark (VLQ) $Q$ is of Nelson-Barr type (NB-VLQ) when its Lagrangian is
\begin{equation}
\label{lag:Q:NB}
-\lag = \bar{q}_{iL} \cY^{d}_{ij} H d_{jR} + \bar{q}_{iL} \cY^{u}_{ij} \tilde{H} u_{jR} 
+\bar{q}_{iL} \cM^{qQ}_{i} Q_{R} + \bar{Q}_L\cM_Q Q_R + h.c.,
\end{equation}
where $d_{jR},u_{jR}$ are the quark singlets, $\cY^d,\cY^u,\cM_Q$ are {\emph{real}} parameters whereas $\cM^{qQ}$ is complex in the same field basis with vanishing $\theta$ angle. The former framework is realized assuming that the fundamental theory respects CP invariance as well as a $\ZZ_2$ symmetry under which $Q$ is odd, and that CP$\times\ZZ_2$ is softly broken only by $\cM^{qQ}$.

The Lagrangian \eqref{lag:Q:NB} depends on $1(\cM_Q)+3(\cY^u)+6(\cY^d)+5(\cM^{qQ})=15$ parameters, as can be seen in the basis in which $\cY^u=\hat{\cY}^u$ is diagonal and $\cY^d=O_{d_L}\hat{\cY}^d$, where $O_{d_L}$ and $\hat{\cY}^d$ are a real orthogonal matrix and a diagonal matrix with real positive eigenvalues, respectively. As we will see, non-trivial correlations among the Lagrangian parameters are necessary to reproduce the flavor and CP-violating structure of the SM below the mass scale of the exotic fermion. The origin of these correlations will not be investigated here; we will simply assume that the necessary structure is generated by some more fundamental physics.

In order to derive the EFT below the mass of the VLQ it is convenient to change field basis. Performing a special unitary rotation among the fields $(q_{iL},Q_L)$ one can re-express \eqref{lag:Q:NB} in the form 
\eqali{
\label{lag:Q:gen}
-\lag &= \bar{q}_{iL} Y^{d}_{ij} H d_{jR} + \bar{q}_{iL} Y^{u}_{ij} \tilde{H} u_{jR} 
\cr
&\quad +\bar{Q}_{L} Y^{Qd}_{j} H d_{jR} + \bar{Q}_{L} Y^{Qu}_{j} \tilde{H} u_{jR} 
+\bar{Q}_LM_Q Q_R
 +h.c.\,.
}
The explicit mapping is found to be
\subeqali[real>gen]{
Y^d&= (\id_3-ww^\dag)^{1/2}\cY^d\,,\quad Y^{Qd}= w^\dag \cY^d \,,
\\
Y^u&= (\id_3-ww^\dag)^{1/2}\cY^u\,,\quad Y^{Qu}= w^\dag \cY^u \,,
\\
\label{WR:MB}
% M^Q&=(\id_n-w^\dag w)^{-1/2}\cM^Q\,.
M^Q&=(1-|w|^2)^{-1/2}\cM^Q\,,
}
where the 3-vector $w$ is implicitly defined by
\eq{
\label{def:w}
w=\cM^{qQ}{M_Q}^{-1}\,,
}
and has a norm in the range
\eq{
0< |w|<1\,.
}
See Appendix \ref{ap:change} for more details. The utility of this new basis is evidenced by the fact that in this basis the heavy state does not mix with the SM fermions before electroweak symmetry breaking, so that $Y^{u,d}_{ij}$ can be identified (up to radiative corrections) as the SM Yukawas and $M_Q$ as the dominant source of VLQ mass. We will call the field basis in \eqref{lag:Q:gen} the ``VLQ basis", as opposed to the ``Nelson-Barr basis" of \eqref{lag:Q:NB}. 

Note that a generic model for a VLQ $Q$ can always be put in the form \eqref{lag:Q:gen}. The key difference is that a 
generic model depends on $1(M^Q)+3(Y^u)+7(Y^d)+6(Y^{Qd})+5(Y^{Qu})=22$ parameters\footnote{This can be seen choosing for 
instance the basis in which $Y^u=\hY^u$ is diagonal, $Y^d=V\hY^d$, and one phase in $Y^{Qu}$ is removed by rephasing 
$Q_L,Q_R$.}, namely 12 parameters in addition to the SM, whereas our model, which originates from \eqref{lag:Q:NB}, only 
has 5 extra parameters, the same number as in a \emph{singlet} NB-VLQ model.

%%%%%%%%%%%%%%%%%%%%%%%%%%%%%%%%%%%%%%%%%%%
\section{Reproducing the SM Yukawas and the CKM matrix}
\label{sec:SM}

In this section we discuss under which conditions the parameters in  \eqref{lag:Q:NB} can reproduce the SM below the scale $M_Q$. As will become evident shortly, this turns out to be possible only in the regime $|w|\sim1$. In that limit there is no small expansion parameter one can use to parametrize CP violation. Furthermore, the matrix $\id_3-ww^\dag$ becomes approximately rank-2, indicating that some correlation among the entries in $\cY^d, \cY^u$ is necessary to recover the quark mass hierarchy. These considerations indicate that the SM flavor structure in NB-VLQ models can be studied reliably only via a numerical analysis. This is what we will do next.

To start, we observe that
\eqali{
\label{Yd.Yu}
(\id_3-ww^\dag)^{1/2}\cY^d{\cY^d}^\tp (\id_3-ww^\dag)^{1/2}&=Y^d{Y^d}^\dag\,,
\cr
(\id_3-ww^\dag)^{1/2}\cY^u{\cY^u}^\tp (\id_3-ww^\dag)^{1/2}&=Y^u{Y^u}^\dag\,.
}
The right-hand side of these relations can be diagonalized via independent rotations
\eq{
\label{YY:yi}
Y^u{Y^u}^\dag=V_{u_L}\diag(y^2_{u_i})V^\dag_{u_L}
\,,\quad
Y^d{Y^d}^\dag=V_{d_L}\diag(y^2_{d_i})V^\dag_{d_L}\,,
}
where $(y_{u_i})=(y_u,y_c,y_t)$ correspond to the eigenvalues of the up quark Yukawas and are determined 
experimentally. Analogous considerations can be made for $y_{d_i}$. However, the matrices $V_{u_L},V_{d_L}$ cannot be 
determined unambiguously. Only the CKM matrix $V$ is constrained by data:
\eq{
\label{ckm}
V_{u_L}^\dag V_{d_L}=V\,.
}
We therefore see that our parameters $w, \cY^d, \cY^u$ are constrained by \eqref{Yd.Yu}, but unfortunately cannot be 
fully determined. In the following subsections we will attempt to extract as much information as possible from 
\eqref{Yd.Yu}. In Subsections \ref{sec:Yukinputs} and \ref{sec:CKMinputs}, respectively, we will see how the quark 
masses and the CKM matrix constrain $w, \cY^d, \cY^u$. From these results the structure of the couplings $Y^{Qu,Qd}$ of 
\eqref{lag:Q:gen} will follow. The latter will be discussed in Subsection \ref{sec:hierarchies}.

%%%%%%%%%%%%%%%%%%%%%%%%%%%%%%%%%%%%%%%%%%%
\subsection{SM Yukawas as input}
\label{sec:Yukinputs}

The relation between the SM $Y$ and the CP conserving $\cY$ in both up and down sectors is given in 
\eqref{real>gen}:
\eq{
\Omega^{1/2}\cY=Y\,,
}
where we introduced the shorthand 
\eq{
\Omega\equiv \id-ww^\dag.
}
The expression $YY^\dag$ gives the relations \eqref{Yd.Yu}, whose mismatch in the diagonalization matrices leads to the CKM mixing. If we take the other hermitian combination, we obtain
\eq{
\label{HR}
\cY^\tp\Omega\cY=Y^\dag Y\,.
}
The procedure to solve this relation for $\cY$, given $Y^\dag Y$ and $w$, was discussed in 
Ref.\,\cite{consequences} for a rank one $w$ and in Ref.\,\cite{nbvlq:more} for $w$ with rank
two or greater.
Applying that procedure to the present scenario allows us to use the SM Yukawas as input 
parameters to constrain $w, \cY^d, \cY^u$. We summarize the procedure in the following and relegate the details to 
Appendix \ref{ap:inversion}.

We first split the right hand side of \eqref{HR} as
\eq{
\label{HR:Z}
Y^\dag Y=\re(Y^\dag Y)+i\im(Y^\dag Y)=X^{1/2}[\id_3+iZ]X^{1/2}\,,
}
where $X=\re(Y^\dag Y)$ is positive definite and $X^{1/2}$ is its square root.
Note that $X$ is real symmetric while $\im(Y^\dag Y)$, and $Z$, are real antisymmetric.
Therefore $Z$ can be parametrized in terms of a (pseudo) vector $\bz=(z_k)$ as
\eq{
\label{def:Z}
Z_{ij}=\eps_{ijk}z_k\,.
}
Although there are two of these vectors, $\bz^u,\bz^d$, one for the up and one for the down sectors, their norm turn out to be the same (see Appendix \ref{ap:inversion}):
\eq{
\label{muu=mud}
\bz^u\cdot \bz^u=\bz^u\cdot \bz^u=\mu^2\,,
}
and entirely determined by $w$. The quantity $\mu$ is a measure of CP violation in ${Y^u}^\dag Y^u, {Y^d}^\dag Y^d$. It is related to 
$w$ as we will see below. 

Performing an orthogonal basis transformation on $d_R$ and $u_R$ will not affect the SM input in \eqref{Yd.Yu}. We can choose a basis where $X$ is diagonal,
\eq{
\label{X:xi}
X=\re(Y^\dag Y)=\diag(x_1,x_2,x_3)\,.
}
This can be achieved independently for $Y^u$ and $Y^d$. The eigenvalues $x_i$ scale as $y_i^2$, with $y_i$ the SM Yukawas.

\begin{figure}[h]
\includegraphics[scale=0.85]{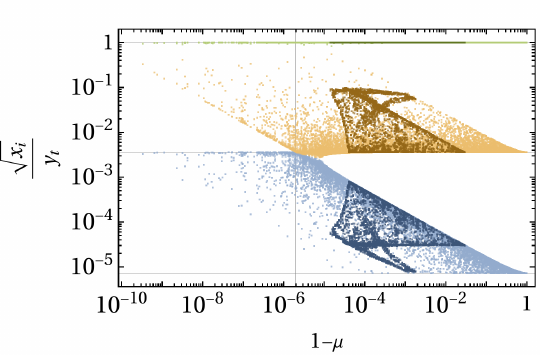}
\includegraphics[scale=0.85]{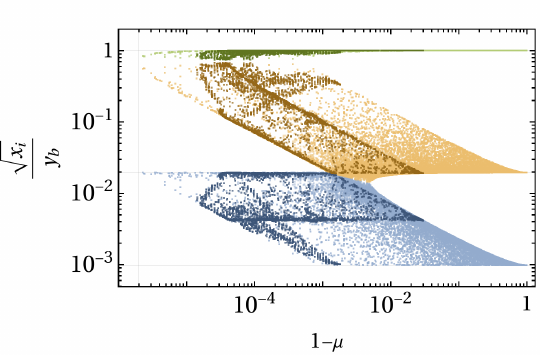}
\caption{\label{fig:x_i-mu}%
Distribution of $\sqrt{x_i}/y_3$ against $1-\mu$ for up (left) and down (right) sectors with $i=3,2,1$ corresponding to 
green, orange and blue points.
Points shown in a lighter shade only reproduce the eigenvalues of the SM Yukawas while the points in a darker shade also 
reproduce the CKM angles and phase. The horizontal lines show $y_i/y_3$ while the vertical line refers to $(1-\mu)_{\rm 
min}$ in 
\eqref{doublet:mu-constraint}.
}
\end{figure}

Now, for a given $\bz$, \eqref{HR:Z} can be solved for $x_i$ in each sector in terms of the eigenvalues $y_i^2$ of $Y^\dagger Y$. One can check that $x_i$ inherit the hierarchy from $y_i^2$, and we find
\eq{
\label{xi:ranges}
y_1^2<x_1<y_2^2\,,\quad
y_2^2\lesssim x_2<y_3^2\,,\quad
x_3\sim y_3^2\,.
}
This can be seen in the points shown in a lighter shade in Fig.\,\ref{fig:x_i-mu}, where we see the possible values for $\sqrt{x_i}/y_3$ 
as a function of $1-\mu$ for each sector. It corresponds to the spectrum of $X^{1/2}/y_3$ while the gray horizontal lines show the 
values of $y_i/y_3$ corresponding to the spectrum of $(YY^\dag)^{1/2}/y_3$.
Qualitatively, the eigenvalues of $X$ lie between the largest and smallest eigenvalues of $YY^\dag$.
The gray vertical line in both panels shows the minimum value of $1-\mu$ that allows solutions for 
\eqref{HR:Z} in the down sector;
see appendix \ref{ap:max:mu}.
For the up sector separately, $1-\mu$ can in principle reach smaller values; but these are discarded because of \eqref{muu=mud}.
For definiteness, we use running SM Yukawas at 1 TeV\,\cite{antusch}.\footnote{More up to date values can be obtained in \cite{Huang:2020hdv} but they do not lead to significant differences.}

Our next goal is to see how these considerations can be used to constrain $\cY$. With this in mind, we go back to the 
left hand side of \eqref{HR} and write
\eq{
\Omega=\id_3-ww^\dag=\Omega_1+i\Omega_2\,.
}
Then the real and imaginary parts of \eqref{HR} become
\eqali{
\label{inversion:re.im}
\cY^\tp\Omega_1\cY&=X\,,
\cr
\cY^\tp\Omega_2\cY&=X^{1/2}ZX^{1/2}\,.
}
These relations can next be solved for $\cY$ in terms of $w$, $x_i$ and $\bz$ which in turn depend on the SM Yukawas 
$y_i$ 
($y_{u_i}$ or $y_{d_i}$) as explained above.
The solution for the real part yields
\eq{
\label{ycal=X}
\cY^\tp =X^{1/2}\cO\Omega_1^{-1/2}\,
\text{~~or~~}
\cY =\Omega_1^{-1/2}\cO^\tp X^{1/2}\,,
}
where $\cO$ is an orthogonal matrix determined by the vector $z$ (see Appendix \ref{ap:inversion}).

Choosing the basis where
\eq{
\label{w:ab}
w=(a,ib,0)^\tp\,,
}
the results of Appendix \ref{ap:inversion} give \cite{nbvlq:more}
\eq{
\label{mu=a.b}
\mu=\frac{ab}{\sqrt{1-a^2}\sqrt{1-b^2}}\,,
}
which is the advertised relation between $\mu$ and $w$. We will interpret this as the relation determining $a$ given the 
pair $(b,\mu)$. Also, observe that nonvanishing $\mu$ implies 
nonvanishing $ab$. The parameter $\mu$ can be also related directly to $Y^\dag Y$; see \eqref{mu:det}.

Let us summarize what we have achieved so far. We started from the 14 parameters $w, \cY^d, \cY^u$ and constrained them 
via \eqref{Yd.Yu} (the 15th parameter is the mass scale $\cM^Q$ and remains undetermined). In this section we showed how 
we can trade them for $\mu,b$ (2 unknown parameters), the SM Yukawas $y_{u_i,d_i}$ (6 measured quantities), and the 
matrices $\cO_{u,d}$ (6 unknown parameters in total). Of the original 14 unknown parameters, 6 have thus been 
determined. Of the remaining 8, namely
\eq{
\label{parameters}
\mu,b,\cO_{u,d}\,,
\,
}
$\mu$ is forced to be in the range shown in Fig.\,\ref{fig:x_i-mu} whereas $b$ and $\cO_{u,d}$ are still unconstrained. 
In the following we 
will see how the CKM matrix can be used to constrain them.

%%%%%%%%%%%%%%%%%%%%%%%%%%%%%%%%%%%%%%%%%%%
\subsection{The CKM as input}
\label{sec:CKMinputs}

We rewrite the relations \eqref{Yd.Yu} using \eqref{ycal=X} and obtain
\eqali{
\label{YYdag:param}
\Omega^{1/2}\Omega_1^{-1/2}\cO_u^\tp X^u\cO_u\Omega_1^{-1/2}\Omega^{1/2}
&=Y^u{Y^u}^\dag\,,
\cr
\Omega^{1/2}\Omega_1^{-1/2}\cO_d^\tp X^d \cO_d\Omega_1^{-1/2}\Omega^{1/2}
&=Y^d{Y^d}^\dag\,.
}
The unitary matrices that diagonalize them determine the CKM matrix according to \eqref{ckm}.

To find the physically relevant parameters we perform a scan in the 8-dimensional space
\eqref{parameters} 
and select only the points that lead to the correct CKM matrix. We allow a $3\sigma$ variation of the CKM parameters at 
1 TeV\,\cite{antusch}. Our numerical analysis is complemented with an analytic one, which makes use of some flavor 
invariants as described in App.\,\ref{sec:invs}.

\begin{figure}[h]
\includegraphics[scale=0.8]{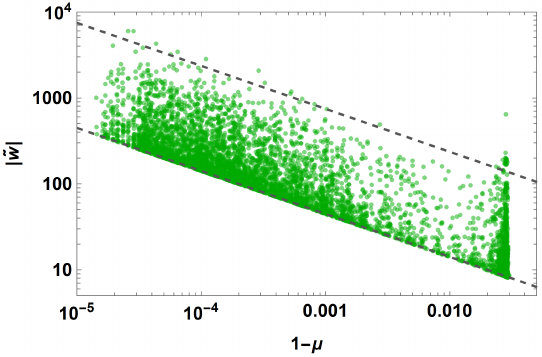}
\caption{\label{fig:deltamu.wtilde}
$|\tw|$ as a function of $1-\mu$ for physical points that reproduce the SM.
The dashed lines obey $|\tw|=\sqrt{\frac{2}{1-\mu}}$ and $|\tw|=\sqrt{\frac{556}{1-\mu}}$, corresponding to 
$\sin\theta=1$ and $0.03$ in \eqref{deltamu:w-tide:approx} approximately.
}
\end{figure}

Some viable points are shown in Fig.\,\ref{fig:x_i-mu} in the plane 
$(1-\mu,\sqrt{x_i}/y_3)$ for both up (left panel) and down (right panel) sectors. The pale-colored points are compatible 
with 
the SM Yukawas, but are not necessarily capable of reproducing the CKM matrix. The darker-colored points 
instead reproduce both the SM Yukawas as well as the CKM matrix. Note that to reproduce the CKM matrix we find that the 
parameter $\mu$ must be rather close to unity:
\eq{
\label{range:1-mu}
1.4\times 10^{-5}\lesssim 1-\mu\lesssim 0.03\,.
}
In this regime, and writing
\eq{
\frac{w}{|w|}=(\cos\theta,i\sin\theta,0)\,,
\quad \theta\in (0,\pi/2)
\,,
}
Eq. \eqref{mu=a.b} can be approximated as
\eq{
\label{deltamu:w-tide:approx}
1-\mu\approx \frac{1-|w|^2}{2s_\theta^2 c_\theta^2}
=\frac{1}{2s_\theta^2 c_\theta^2(1+|\tw|^2)}
\,.
}
This implies that 
\eq{
\label{w=1}
|w|\approx 1.
}
We show in Fig.\,\ref{fig:deltamu.wtilde} the quantity (see \eqref{tw:cM})
\eq{
\label{|tw|}
\tw=\cM^{qQ}/{\cM^Q}=\frac{w}{\sqrt{1-|w|^2}}
}
as a function of $1-\mu$ and see that 
\eqref{deltamu:w-tide:approx} is valid to a good approximation. Our scan only covers the domain $\sin\theta\gtrsim 
0.03$.

Eq. \eqref{w=1} is the relation we alluded to at the beginning of Section \ref{sec:SM}. It is a necessary condition to 
reproduce the CKM structure, and also the technical reason why our numerical analysis is necessary to identify in which 
regime \eqref{lag:Q:NB} reproduces the SM at low energies.

%%%%%%%%%%%%%%%%%%%%%%%%%%%%%%%%%%%%%%%%%%%
\subsection{Hierarchies in the VLQ couplings}
\label{sec:hierarchies}

Having identified the relevant region of parameter space, we can now discuss the resulting flavor structure of the 
couplings $Y^{Qu,Qd}$ of the VLQ doublet to the SM quarks, as defined in the VLQ basis \eqref{lag:Q:gen}.

Using the convenient notation \eqref{|tw|} (see Appendix \ref{ap:change}) we rewrite the relations \eqref{real>gen} in 
the basis where $Y^u=\hY^u$ is diagonal:
\eq{
\label{YQu:YQd:SM}
Y^{Qu}=\tw^\dag \hY^u\,,\quad
Y^{Qd}=\tw^\dag V \hY^d\,;
}
with $V$ the CKM matrix.
It is clear that if all the components of $\tw=(\tw_1,\tw_2,\tw_3)$ are of the same order, these couplings inherit the 
hierarchies of the SM Yukawas.
In contrast, the overall scale of the couplings are not determined by the SM Yukawas. It is controlled 
by $|\tw|$ in \eqref{|tw|} which, as seen in Fig.\,\ref{fig:deltamu.wtilde}, can be significantly larger than unity for 
the doublet NB-VLQ (similarly to the singlet versions \,\cite{nbvlq:more}).

\begin{figure}[h]
\includegraphics[scale=0.8]{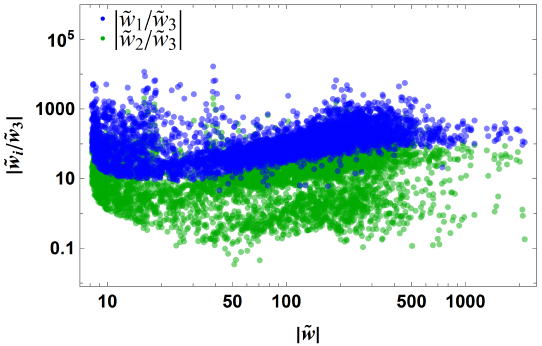}
\caption{\label{fig:wtilde-ratios}
Ratios $|\tw_i|/|\tw_3|$, $i=1,2$ against the norm $|\tw|$. All points reproduce the SM.
}
\end{figure}

We first inspect the relative size of the components $\tw_i$, working in the basis $Y^u=\hY^u$ (rather than the one we 
adopted in the parametrization \eqref{w:ab}).
We show in Fig.\,\ref{fig:wtilde-ratios} the ratios $|\tw_1/\tw_3|$ and $|\tw_2/\tw_3|$ as a function of $|\tw|$.
We see that within a broad variation, there is an \emph{inverted} hierarchy which roughly follows
\eq{
\label{wtilde:hierarchy}
|\tw_1|:|\tw_2|:|\tw_3|\approx 200:10:1\,.
}
The inverted hierarchy for $\tw_i$ imply that the components $Y^{Qu}_i$ and $Y^{Qd}_i$ exhibit a hierarchy less 
pronounced than the corresponding Yukawa couplings of the SM. From the rough estimate \eqref{wtilde:hierarchy} indeed we 
expect the components of \eqref{YQu:YQd:SM} to scale as
\eqali{
\label{estimate:YQ}
|Y^{Qu}_i|&\sim |\tw|(y_u,\ums{20}y_c,\ums{200}y_t)
\sim |\tw|(6\times 10^{-6},1.5\times 10^{-4},4\times 10^{-3})\,,
\cr
|Y^{Qd}_i|&\sim |\tw|(y_d,0.22 y_s, 0.01 y_b)
\sim 
|\tw|(10^{-5},6\times 10^{-5},10^{-4})\,.
}
where in the estimation for $Y^{Qd}$ we used
\eq{
\tw^\dag V\sim |\tw|(1,|V_{ud}|,|V_{ub}|+\ums{20}|V_{ub}|+\ums{200}|V_{tb}|)
\sim |\tw|(1,0.22,0.01)\,.
}
Note that the hierarchies in \eqref{estimate:YQ} differ from the case of singlet NB-VLQs of up or down type which 
roughly follow the hierarchies of the CKM third row or column, respectively\,\cite{nb-vlq:fit}.

\begin{figure}[h]
\includegraphics[scale=0.8]{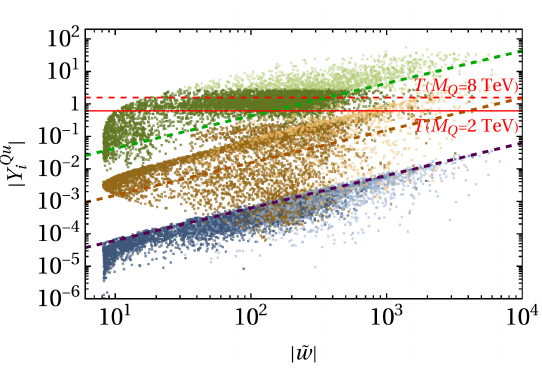}
\includegraphics[scale=0.8]{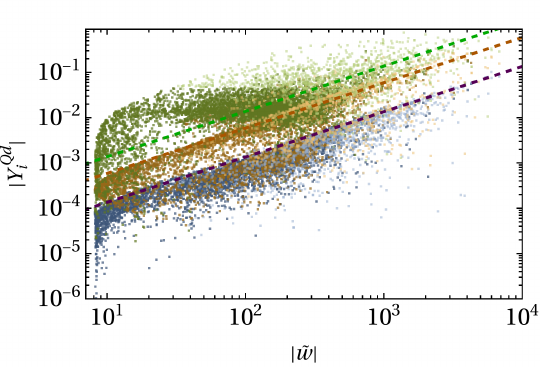}
\caption{\label{fig:YQ.wtilde}%
$|Y^{Qu}_i|$ (left) and $|Y^{Qd}_i|$ (right) as a function of $|\tw|$
for $i=3,2,1,$ respectively in green, orange and blue.  
The lighter green points are excluded by the perturbativity 
limit \eqref{cYu<3}.
The dashed lines with the same color as the points show the approximate functions in \eqref{estimate:YQ}.
The horizontal continuous (dashed) red line indicate the constraint coming from electroweak precision observables for 
$M_Q=2\,\unit{TeV}$ ($M_Q=8\,\unit{TeV}$); see Sec.\,\ref{sec:S.T}.
}
\label{YQq}
\end{figure}

\begin{figure}[h]
\includegraphics[scale=0.8]{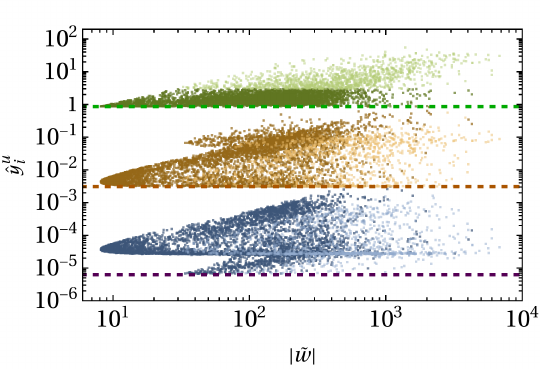}
\includegraphics[scale=0.8]{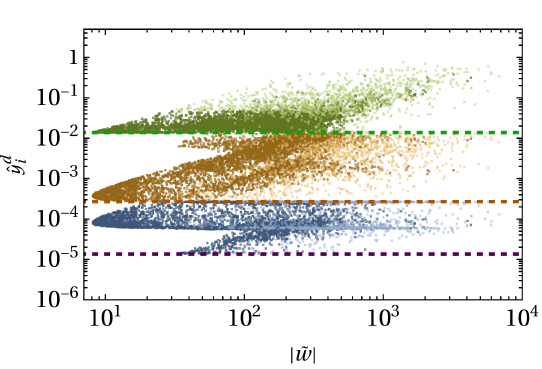}
\caption{\label{fig:calY}%
Eigenvalues $\hcY^u_i$ (left) and $\hcY^d_i$ (right) of the CP conserving Yukawa couplings as functions of $|\tw|$.
The dashed lines correspond to the values of the SM Yukawas at 1\,TeV. The color coding is the same as in 
Fig.\,\ref{fig:YQ.wtilde}.
}
\label{calY}
\end{figure}

The qualitative hierarchy in the Yukawas is confirmed by the more accurate numerical analysis shown in 
Fig.\,\ref{fig:YQ.wtilde}. The colored dots show the actual numerical values of $|Y^{Qu}_i|$ and $|Y^{Qd}_i|$ 
against $|\tw|$ whereas the rough estimates \eqref{estimate:YQ} are shown in the dashed lines.
We note that the Yukawa couplings can actually vary over more than one order of magnitude compared to our crude estimate 
\eqref{estimate:YQ}. For comparison, we also show in Fig.\,\ref{fig:calY} the behavior of the eigenvalues of the CP 
conserving Yukawas $\cY$ against $|\tw|$. They are always larger than the SM values, as we also emphasize in Sec. 
\ref{sec:pert}.

\begin{figure}[h!]
\includegraphics[scale=0.8]{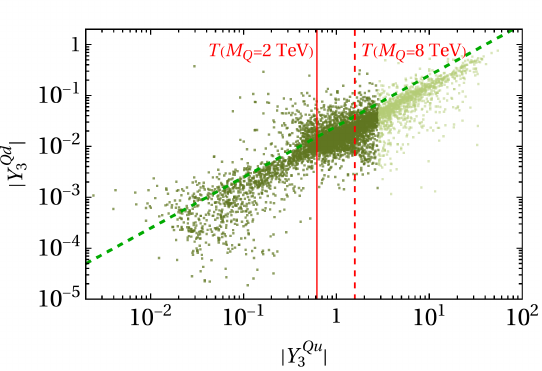}
\caption{\label{fig:YQ3:u.d}%
Correlation between $|Y^{Qu}_3|$ and $|Y^{Qd}_3|$.
The darker and lighter green points are as explained in Fig.\,\ref{fig:YQ.wtilde}.
The green dashed line shows the estimate \eqref{estimate:YQ}.
}
\end{figure}

Finally, Fig.\,\ref{fig:YQ3:u.d} reveals a strong positive correlation between $|Y^{Qu}_3|$ and $|Y^{Qd}_3|$, and of 
course that $|Y^{Qu}_3|\gg |Y^{Qd}_3|$.
A similar correlation is found between $|Y^{Qu}_i|$ and $|Y^{Qd}_i|$ as suggested by the estimate \eqref{estimate:YQ}.
These features qualitatively distinguish scenarios with NB-VLQs from generic doublet VLQs.

%%%%%%%%%%%%%%%%%%%%%%%%%%%%%%%%%%%%%%%%%%%%%%%%%%%%%%%%%%%%%%%%%%%%%%%%%%%%%%%%%%%%

%%%%%%%%%%%%%%%%%%%%%%%%%%%%%%%%%%%%%%%%%%%%%%%%%%%%%%%%%%%%%%%%%%%%%%%%%%%%%%%%%%%%%%
\section{Phenomenological constraints}
\label{sec:constraints}

Here we review the most relevant constraints on our models. These include collider constraints, perturbativity bounds, 
precision electroweak constraints and flavor constraints. Constraints from hadronic CP violation will be discussed in 
Section 
\ref{sec:invariants}.

First, from direct collider searches\,\cite{Benbrik:2024fku} the VLQ must be heavier than
\eq{
\label{MQ:collider}
M_Q>1.5\,\unit{TeV}\,,
}
for a VLQ that couples only to the third SM family. This is approximately the case for NB-VLQs.
For a review on the constraints for generic singlet VLQs, see also Ref.\,\cite{Alves:2023ufm}.

Let us next discuss the other bounds in turn.

\subsection{Perturbativity} 
\label{sec:pert}

In order for our models to be predictable, the new couplings must all be small to allow a perturbative expansion. In the 
basis \eqref{lag:Q:NB} the new couplings are $\cY$, whereas in the VLQ basis they are $Y^{Qu,Qd}$. Using the relation in 
\eqref{real>gen} we find that
\eq{
|Y^{Qu}|^2\le \tr[\cY^u{\cY^u}^\tp]|w|^2
< \sum_i(\hcY^u_i)^2
\approx (\hcY^u_3)^2\,,
}
where the first inequality follows from Schwartz inequality and the second one from $|w|<1$. We thus see that the 
perturbativity constraint on $\cY^u$ is generally stronger. Constraints arising from $\cY^d$ are obviously less relevant 
because the down-type Yukawas are smaller; we will therefore restrict our attention to the up quark sector.

From the relation \eqref{real>gen} or \eqref{HR} between $\cY^u$ and $Y^u$, we also see that the largest eigenvalue 
$\hcY^u_3$ of $\cY^u$ is always larger than the largest eigenvalue $y_t$ of $Y^u$, i.e., the top Yukawa. This is 
confirmed by our numerical analysis shown in Fig.\,\ref{fig:calY}. An upper bound on $\hcY^u_3$ can therefore represent 
an important constraint for our theory, as $y_t$ is already close to unity in the SM.

To provide a rough upper bound on $\hcY^u_3$ we study its RG evolution and require that a hypothetical Landau pole be 
sufficiently heavier than the scale $M_Q$ of the VLQ. From the Lagrangian \eqref{lag:Q:NB} we see that the running of 
$\hcY^u_3$ at one-loop is governed by the same RGE as the top Yukawa in the SM\,\cite{Buttazzo:2013uya}. Retaining only 
the dominant contributions we have
\eq{
16\pi^2\mu\frac{d\hcY^u_3}{d\mu}=\frac{9}{4}(\hcY^u_3)^3-4g_s^2\hcY^u_3\,,
}
which reveals that the beta function is positive whenever $\hcY_3>4g_s/3 \gtrsim 1.3$.
Starting at the scale $\mu=M_Q$ and requiring that the Landau pole stays above $10M_Q$ one arrives at
\eq{
\label{cYu<3}
\hcY^u_3(M_Q)\lesssim 3\,.
}
This requirement is significantly more constraining than imposing a naive perturbativity bound $\hcY^u_3<4\pi$. We will 
therefore adopt \eqref{cYu<3} in the following.

\subsection{Electroweak precision observables} 
\label{sec:S.T}
 
The vector-like quark doublet induces loop corrections to electroweak precision observables which can be 
parameterized by the oblique parameters $S$ and $T$. The expressions, recalling that $Y^{Qu}_3,Y^{Qd}_3$ dominate over 
the other Yukawa couplings, are\,\cite{Belfatto:2023tbv,Chen:2017hak}
\subeqali[S.T:vlq]{
\Delta S&=S-S_{\text{SM}}
\cr
&=\frac{3}{18 \pi}\frac{v^2}{M^2_Q}\left(|Y^{Qu}_3|^2\left[-10+4 \log \left(\frac{M_{Q}^{2}}{m_{t}^{2}}\right)\right]
+|Y^{Qd}_3|^2\left[-6+2 \log \left(\frac{M_{Q}^{2}}{m_{b}^{2}}\right)\right]\right)\,,
\\
\Delta T&= T-T_{\text{SM}}
\cr
&=\frac{3v^2 }{8 \pi s_{W}^{2} M_{W}^{2}}\left(\frac{m_{t}^{2}|Y^{Qu}_3|^2}{M^2_Q}\left[-3+2 \log 
\left(\frac{M_{Q}^{2}}{m_{t}^{2}}\right)\right]+\frac{2}{3}\frac{v^2}{M^2_Q}\left(|Y^{Qu}_3|^2-|Y^{Qd}_3|^2
\right)^2\right)
\,,
}
where $v=246$ GeV. The parameters $S,T$ (for $\Delta U=0$) are currently 
constrained at 95\% CL to be within\,\cite{ParticleDataGroup:2024cfk}, 
\eqali{
\label{S.T:pdg}
\Delta S=-0.05\pm 0.07, \quad  \Delta T=0.00\pm 0.06\,,
}
with a correlation $\rho=0.92$.

In our model $|\Delta T|\gg |\Delta S|$. We can therefore neglect $\Delta S$. The constraints arising from $\Delta T$ 
are shown in Fig.\,\ref{fig:YQ.wtilde} as a horizontal solid line for $M_Q=2$ TeV and dashed line for $M_Q=8$ TeV. To 
get a qualitative understanding, we note that for $Y^{Qu}_3\lesssim1.7$ the $Y^2y_t^2/M^2$ contribution to $\Delta T$ 
dominates and we estimate the 2-$\sigma$ bound 
\eq{
\label{T:rough}
\frac{M_Q}{\unit{TeV}}\gtrsim2.5\,|Y^{Qu}_3|\sim10^{-2}|\tw|
\,,
}
where in the last relation we used the rough approximation \eqref{estimate:YQ}. For $Y^{Qu}_3\gtrsim3$ the $Y^4/M^2$ 
term dominates, but that limit is incompatible with our perturbative constraint \eqref{cYu<3}. In the interesting regime 
$1.7\lesssim Y^{Qu}_3\lesssim3$ both terms contribute comparably. In general, for a fixed $Y^{Qu}_3$ the corrections to 
$\Delta T$ decouple as $M_Q\gg1$ TeV. This can be clearly seen in Fig.\,\ref{fig:YQ.wtilde} by comparing the horizontal 
solid gray line at $M_Q=2\,\unit{TeV}$ with the dashed one at  
$M_Q=8\,\unit{TeV}$. With $M_Q\sim 12\,\unit{TeV}$ the constraint obtained from the $T$ parameter becomes comparable to 
the perturbativity limit of \eqref{cYu<3}.

%%%%%%%%%%%%%%%%%%%%%%%%%%%%%%%%%%%%%%%%%%%
\subsection{Flavour constraints} 
\label{sec:FV}

Generic models of VLQs are significantly constrained by flavor observables. In NB-VLQ scenarios, instead, the Yukawa 
couplings in \eqref{lag:Q:gen} have naturally a hierarchical structure, as seen in Fig. \ref{YQq}. As a consequence, we 
will argue that flavor constraints are less relevant than the perturbativity bound and the electroweak precision 
observables discussed earlier.

Doublet VLQs induce the SMEFT operators $O_{Hu}, O_{Hd}, O_{Hud}$ ($\psi^2 H^2D$) and $O_{uH}, 
O_{dH}$ ($\psi^2H^3$) at tree-level\,\cite{delAguila:2000rc}. We first discuss flavor violation and subsequently 
flavor-diagonal transitions.

The operators $O_{Hu}, O_{Hd}$ lead to flavour-changing $Z$ couplings of the right-handed quarks, which induce primarily 
dangerous $\Delta F=1$ transitions\,\cite{Ishiwata:2015cga,Bobeth:2016llm}. The limits coming from $\Delta F=1$ 
transitions $d_i\to d_j$ and $u\to c$ are\,\cite{Ishiwata:2015cga}
\eqali{
\label{flavor:DeltaF=1}
\frac{M^2_Q}{\unit{TeV}^2} &> 66^2\times |Y_1^{{Qd}^*} Y_2^{Qd}|\sim (1.6\times 10^{-3}|\tw|)^2\,,
\cr
\frac{M^2_Q}{\unit{TeV}^2}&> 100^2\times |\text{Re}(Y_1^{{Qd}^*} Y_2^{Qd})|\sim (2.4\times 10^{-3}|\tw|)^2\,,
\cr
\frac{M^2_Q}{\unit{TeV}^2}&> 30^2\times |Y_3^{{Qd}^*} Y_1^{Qd}|\sim (9.5\times 10^{-4}|\tw|)^2\,,
\cr
\frac{M^2_Q}{\unit{TeV}^2}&> 18^2\times |Y_3^{{Qd}^*} Y_2^{Qd}|\sim (1.4\times 10^{-3}|\tw|)^2\,,
\cr
\frac{M^2_Q}{\unit{TeV}^2}&> 3.9^2\times |Y_1^{{Qu}^*} Y_2^{Qu}|\sim (1.2\times 10^{-4}|\tw|)^2\,,
}
where the last approximation uses the estimate \eqref{estimate:YQ}. Consider for instance the {fourth} 
constraint, coming from $B_s\to \mu^+\mu^-$, which 
leads to $M_Q/\unit{TeV}\gtrsim 1.4\times 10^{-3}|\tw|$. This bound, like all the others quoted in 
\eqref{flavor:DeltaF=1}, is weaker than the rough lower limit from $\Delta T$ in \eqref{T:rough}.

The operators $O_{Hu}, O_{Hd}$ also contribute to $\Delta F=2$ transitions through $Z$ exchange. However, for $M_Q$ 
above $\sim1$ TeV such processes are dominated by direct one-loop contributions from box 
diagrams\,\cite{Bobeth:2016llm}. Actually, for the $(sd)$ sector, the latter are further superseded by the 
contributions from $\bar{L}L\bar{R}R$ four-fermion operators at the electroweak scale induced by one-loop RG mixing 
from 
the tree-level generated $O_{Hd}$. Flavor-violating transitions mediated by $O_{uH}, O_{dH}$ are suppressed by the SM 
Yukawas and can be neglected.
In summary, the most relevant $\Delta F=2$ bounds are\footnote{%
It seems to us that the $\Delta F=2$ bounds quoted for the 
couplings $Y^{Qu}_i$ in the ($tu$) and ($tc$) sectors and the doublet $Q$ in Table~1 of Ref.\,\cite{Ishiwata:2015cga} 
(coupling $\lambda^{(u)}_i=Y^{Qu}_i$) might have a typo.}
\eqali{
\label{flavor:DeltaF=2}
\frac{M_Q^2}{\unit{TeV}^2}&> 59^2\times \Big|\text{Re}({Y_1^{Qd}}^* Y_2^{Qd})^2\Big|
\sim \big(1.9\times 10^{-4}|\tw|\big)^4
&&\text{\cite{Ishiwata:2015cga}},
\cr
\frac{M^2_Q}{\unit{TeV}^{2}}&> 10^6\times \Big|\im({Y_1^{Qd}}^*Y_2^{Qd})^2\Big|
\sim \big(7.7\times 10^{-4}|\tw|\big)^4
&&\text{\cite{Glioti:2024hye}},
\cr
\frac{M_Q}{\unit{TeV}}&> 44\times |{Y_3^{Qd}}^* Y_1^{Qd}| 
\sim (2.1\times 10^{-4}|\tw|)^2
&&\text{\cite{Glioti:2024hye}},
\cr
\frac{M_Q}{\unit{TeV}}&> 9.1\times|{Y_3^{Qd}}^* Y_2^{Qd}|
\sim (2.3\times 10^{-4}|\tw|)^2
&&\text{\cite{Ishiwata:2015cga}},
\cr
\frac{M_Q}{\unit{TeV}}& >67\times |{Y_1^{Qu}}^* Y_2^{Qu}|
\sim (2.5\times 10^{-4}|\tw|)^2\,,
&&\text{\cite{Ishiwata:2015cga}},
\cr
\frac{M^2_Q}{\unit{TeV}^{2}}& >400^2\Big|\im({Y_1^{Qu}}^*Y_2^{Qu})^2\Big|
\sim \big(6\times 10^{-4}|\tw|\big)^4
&&\text{\cite{Glioti:2024hye}},
}
where the last approximations follow from \eqref{estimate:YQ}. We note that the most stringent constraints on $|\tw|$ 
are the second and the last ones, corresponding to $\eps_K$ and $D^0$ mixing respectively.
The first is stronger than \eqref{T:rough} only when $|\tw|\gtrsim 1.7\times 10^4$ which is however outside of the 
physical range of $|\tw|$; see Fig.\,\ref{fig:deltamu.wtilde}.

For completeness, we also report the constraint from flavor changing decays $t\to u_iZ$ and $t\to u_iH$ in 
colliders which are comparable to low-energy constraints from flavor observables\,\cite{Belfatto:2023tbv}:
\eqali{
\frac{M_Q^2}{\unit{TeV}^2}\gtrsim 3|Y^{Qu}_1Y^{Qu}_3|\sim (2.7\times 10^{-4}|\tw|)^2\,,
\cr
\frac{M_Q^2}{\unit{TeV}^2}\gtrsim 2.3\times |Y^{Qu}_2Y^{Qu}_3|\sim (1.1\times 10^{-3}|\tw|)^2\,,
\cr
\frac{M_Q^2}{\unit{TeV}^2}\gtrsim 0.37\times y_t|Y^{Qu}_1Y^{Qu}_3|\sim (9\times 10^{-5}|\tw|)^2\,,
\cr
\frac{M_Q^2}{\unit{TeV}^2}\gtrsim 0.32\times y_t|Y^{Qu}_2Y^{Qu}_3|\sim (4.4\times 10^{-4}|\tw|)^2\,,
}
These are also weaker than the $\Delta T$ bound, though they often apply to different combinations of $Y^{Qu}$'s as the 
previous constraints.

It remains to evaluate the impact of the flavor-conserving transitions beyond the SM. The flavor-diagonal components in 
$O_{Hu}, O_{Hd}, O_{uH}, O_{dH}$ are not particularly relevant in our models. The reason is that the couplings to the 
light generations are strongly suppressed by the hierarchy of $Y^{Qu,Qd}$. The operator 
$O_{Hud}=\overline{u}_R\gamma^\mu 
d_R\,\widetilde H^\dagger i{D}_\mu H$, generated in our models at tree-level with Wilson coefficient 
$C_{Hud}={Y^{Qu}}^\dag Y^{Qd}/M_Q^2$, is more interesting. After the Higgs has acquired its vacuum expectation value, 
that operator mediates exotic couplings of the $W^\pm$ to the right-handed quarks which are not present in the SM at 
tree-level. Following the discussion of Ref.\,\cite{Vignaroli:2012si}, these exotic interactions induce a novel 
contribution $\delta C_7(\mu_{ew})=-0.777 (m_t/m_b)[C_{Hud}]_{33}v^2/2$ to the effective dipole operator $O_7$ 
describing $b\to s+\gamma$. Using the updated experimental average of HFLAG\,\cite{hflav.22} with the theoretical SM 
prediction\,\cite{misiak}, we find at 95\%CL
\eq{
-0.038<\re\delta C_7(\mu_{ew})< 0.26\,,
}
which reads
\eq{
\frac{M_Q^2}{\unit{TeV}^2}\gtrsim 5.5^2\times |Y_3^{Qu}Y_3^{Qd}|\sim (3.5\times 10^{-3}|\tw|)^2\,.
}
For a given $|\tw|$ this bound is again weaker than the one in \eqref{T:rough} from $\Delta T$. Yet, it constrains a 
different combination of Yukawas than those collected above.

We should emphasize that the limits on $M_Q^2/Y^2$ and $M_Q^2/Y^4$ quoted in this section, as well as the one in 
\eqref{T:rough}, are accurate, as written. However, the bounds on $M_Q$ shown as a function of $\tw$ are all inherently 
approximate because they are based on the rough estimate \eqref{estimate:YQ}. We checked that the bounds quoted 
here and \eqref{T:rough} are generically conservative. In fact, analogously to the singlet case\,\cite{nbvlq:more}, 
there might be regions in the parameter space where special cancellations take place and the lower bounds on $M_Q$ can 
be further relaxed.

%%%%%%%%%%%%%%%%%%%%%%%%%%%%%%%%%%%%%%%%%%%%%%%%%%%%%%%%%%%%%%%%%%%%%%%%%%%%%%%%%%%%%%
\section{Irreducible contributions to $\btheta$}
\label{sec:invariants}

The main challenge faced by scenarios with spontaneous CP violation is to make sure that corrections to $\bar\theta$ 
remain under control even after spontaneous symmetry breaking. In Nelson-Barr models there are two classes of radiative 
corrections. The first class involves the degrees of freedom of the Higgs sector responsible for CP violation. These 
are 
rather model-dependent, but can be naturally made small if the couplings of the new scalars are sufficiently 
suppressed, as in \,\cite{vecchi.2}. The second class of corrections to $\bar\theta$ arises from loops of the SM and the 
VLQ, 
which effectively communicates CP-violation to the SM. These corrections are unavoidable in these models and are 
independent of how CP is 
broken. They have therefore been dubbed ``irreducible" in \cite{Valenti:2021rdu}. Crucially, corrections to $\bar\theta$ 
do not decouple as the mass scale of the VLQ is increased.

The paper \cite{Valenti:2021rdu} presents a comprehensive discussion of the irreducible corrections to $\bar\theta$ in 
models with electroweak-singlet vector-like fermions, i.e. models with $u$-mediation and $d$-mediation. Scenarios with 
$q$-mediation --- namely the scenarios considered here --- were erroneously dismissed on the grounds of allegedly large 
radiative corrections to $\bar\theta$. That claim originates from the earlier work \cite{vecchi.14}, which provided 
model-independent two-loop estimates applicable (among others) to {\emph{generic}} $q$-mediation frameworks. However, 
what 
\cite{Valenti:2021rdu} overlooked is that, in minimal Nelson-Barr scenarios of $q$-mediation, an accidental symmetry 
prevents such 2-loop corrections to arise.\footnote{LV would like to thank Alessandro Valenti, who crucially 
contributed 
to the realization of this point.} In this subsection we clarify this important aspect and argue that the first 
irreducible corrections to $\bar\theta$ in renormalizable Nelson-Barr models for $q$-mediation (the models discussed in 
this paper) 
arise only at 3-loops, and can be naturally within the current experimental bounds.

The contributions to $\bar\theta$ are more conveniently identified in the basis \eqref{lag:Q:gen} and taking advantage 
of the spurionic flavor symmetries. In that language the coupling $Y^u$ can be seen as a spurion transforming under 
flavor 
$U(3)_q\times U(3)_u\times U(3)_d\times U(n)_Q$ as $Y^u\to U_qY^uU_u^\dagger$, with $U_q\in U(3)_q$ and analogously 
$U_u\in U(3)_u$. Similar considerations reveal that $Y^d\to U_qY^d U_d^\dagger$ whereas $Y^{Qu}\to 
U_QY^{Qu}U_u^\dagger$ 
and $Y^{Qd}\to U_QY^{Qd}U_d^\dagger$. 

Now, recalling that in the SM 
$$
\bar\theta=\theta-{\text{Arg}}[{\text{Det}}(Y^{u,{\text{SM}}}){\text{Det}}(Y^{d,{\text{SM}}})]\,,
$$
we see that the corrections to $\bar\theta$ belong to two distinct classes. The first includes direct contributions to 
$\theta$\,\cite{running.theta}. They are parametrized by polynomial CP-odd, flavor-invariant combinations of the 
couplings of the theory. The 
second class comes from corrections to the Yukawa couplings. The spurionic flavor symmetries and the fact that 
$Y^{Qu,Qd}\propto Y^{u,d}$ (see \eqref{YQuYu} and \eqref{YQdYd} in the appendix) indicate that in our models 
these are proportional to the tree-level Yukawas. Specifically, they are of the form $Y^{u,{\text{SM}}}=F_uY^u$ and 
$Y^{d,{\text{SM}}}=F_dY^d$, where $F_{u,d}=1+\delta F_{u,d}$ with $1$ the tree-level contribution and $\delta F_{u,d}$ 
polynomial 
combinations of the couplings parametrizing the radiative effects. However, 
${\text{Arg}}[{\text{Det}}(Y^{u,{\text{SM}}}){\text{Det}}(Y^{d,{\text{SM}}})]={\text{Arg}}[{\text{Det}}(F_u){\text{Det}}
(F_d)]={\text{Im}}[\ln{\text{Det}}(F_{u})]+{\text{Im}}[\ln{\text{Det}}(F_{d})]$, since NB models by construction do not 
generate corrections at tree-level. So, radiative corrections to the Yukawas are parametrized by structures like 
${\text{Im}}[\ln{\text{Det}}(F_{u,d})]$, which are also polynomial CP-odd flavor-invariant combinations of the 
couplings. An exhaustive analysis of all corrections to $\bar\theta$ can therefore be obtained by studying the 
polynomial flavor-invariants.
For example, in the case of one singlet VLQ of down or up-type, all of the CP odd invariants can be written as a 
linear combination of 9 basic CP odd invariants multiplied by CP even invariants\,\cite{deLima:2024vrn}.

The basic objects necessary to construct invariants are
\eqali{
\label{buildingblocks}
{Y^u}^\dagger Y^u,~~~{Y^d}^\dagger Y^d,~~~{Y^u}^\dagger Y^d,~~~{Y^{Qu}}^\dagger Y^{Qu},~~~{Y^{Qd}}^\dagger 
Y^{Qd},~~~{Y^{Qu}}^\dagger Y^{Qd}.
}
Corrections beyond the SM are of course parametrized by invariants involving the last three building blocks in 
\eqref{buildingblocks}. In addition, since $Y^u, Y^d, Y^{Qu}, Y^{Qd}$ are perturbative, one should look for expressions 
with the lowest number of coupling insertions. The CP-odd flavor-invariant with the lowest number of building blocks 
turns out to be \cite{vecchi.14}
\eqali{
\label{Inv0}
{\text{Im}}\,{\text{Tr}}\left[{Y^u}^\dagger{Y^d}{Y^{Qd}}^\dagger{Y^{Qu}}-{\text{hc}}\right]={\text{Im}}\left[\tw^\dagger
[Y^u{Y^u}^\dagger,Y^d{Y^d}^\dagger]\tw\right].
}
Now, the crucial point is that radiative corrections proportional to this combination and no other couplings are 
possible in the generic theories considered in \cite{vecchi.14} but cannot be 
induced by the theory in \eqref{lag:Q:gen}. Indeed, the latter enjoys an accidental spurious symmetry under which the up 
and down 
quarks are exchanged, $H\to\tilde{H}$, and 
\eqali{
Y^u\leftrightarrow Y^d,~~~Y^{Qu}\leftrightarrow Y^{Qd},~~~{\mathtt Y}_u\leftrightarrow {\mathtt Y}_d,~~~{\mathtt 
Y}_H\to-{\mathtt Y}_H,
}
where ${\mathtt Y}_{u,d}$, ${\mathtt Y}_H$ are the hypercharges of the right-handed quarks and the Higgs doublet, 
respectively. The combination \eqref{Inv1} is obviously odd under that spurious symmetry and therefore cannot be 
generated unless an odd number of ${\mathtt Y}_H$ appears or contributions proportional to more Yukawa are considered. 
In the 
former case one is forced to include at least a loop involving the hypercharge vector. Adding appropriate powers of 
$1/16\pi^2$ for each $\hbar$ factor, we estimate that the largest CP-odd invariant with a single hypercharge loop is
\eqali{
\label{Inv1}
I_1
&=c_1\frac{g'^2}{(16\pi^2)^3}{\mathtt Y}_H({\mathtt Y}_u+{\mathtt 
Y}_d){\text{Im}}\,{\text{Tr}}\left[{Y^u}^\dagger{Y^d}{Y^{Qd}}^\dagger{Y^{Qu}}-{\text{hc}}\right]\\
&=c_1\frac{g'^2}{(16\pi^2)^3}{\mathtt Y}_H({\mathtt Y}_u+{\mathtt 
Y}_d){\text{Im}}\left[\tw^\dagger[Y^u{Y^u}^\dagger,Y^d{Y^d}^\dagger]\tw\right],
}
up to some calculable real number $c_1$. As already anticipated, other potentially important contributions to 
$\bar\theta$ are generated by loops including additional powers of the Yukawa couplings. We find two possibilities with 
a single 
additional $Y$ pair. They are
\eqali{
\label{Inv2}
I_2
&=c_2\frac{1}{(16\pi^2)^3}{\text{Im}}\,{\text{Tr}}\left[{Y^u}^\dagger{Y^d}{Y^{Qd}}^\dagger{Y^{Qu}}{Y^u}^\dagger 
Y^u-{\text{hc}}\right]+(u\leftrightarrow d)\\
&=c_2\frac{1}{(16\pi^2)^3}{\text{Im}}\left[\tw^\dagger[(Y^u{Y^u}^\dagger)^2,Y^d{Y^d}^\dagger]\tw\right]
+(u\leftrightarrow d),
}
and
\eqali{
\label{Inv3}
I_3
&=c_3\frac{1}{(16\pi^2)^3}{\text{Im}}\,{\text{Tr}}\left[{Y^u}^\dagger{Y^d}{Y^{Qd}}^\dagger{Y^{Qu}}{Y^{Qu}}^\dagger 
Y^{Qu}-{\text{hc}}\right]+(u\leftrightarrow d)\\
&=c_3\frac{1}{(16\pi^2)^3}{\text{Im}}\left[\tw^\dagger[Y^u{Y^u}^\dagger,Y^d{Y^d}^\dagger]\tw\right]\left(\tw^\dagger 
Y^u{Y^u}^\dagger\tw-\tw^\dagger Y^d{Y^d}^\dagger\tw\right),
}
where $c_2, c_3$ are again calculable real numbers.

Eqs. \eqref{Inv1}, \eqref{Inv2} and \eqref{Inv3} arise from irreducible 3-loop contributions involving virtual quarks, 
Higgses, gauge-bosons and VLQs. They are expected to parametrize the dominant corrections to $\btheta$ in models of 
$q$-mediation according to $\delta\bar\theta=I_1+I_2+I_3$. Additional contributions to $\btheta$ involve higher powers 
of Yukawa and gauge couplings or decouple as 
the VLQ mass is much higher than the weak scale \cite{Valenti:2021rdu}. In either case they are numerically smaller 
than 
\eqref{Inv1}, \eqref{Inv2} and \eqref{Inv3}.

A careful numerical evaluation of \eqref{Inv1}, \eqref{Inv2} and \eqref{Inv3} --- obtained assuming $c_1=c_2=c_3=1$ --- 
will be discussed in the subsequent section. However, in order to gauge how strongly $q$-mediation scenarios are 
constrained by current data, it is useful to first provide an analytical estimate. To this end, we exploit 
the flavor-invariant nature of our expressions to go in a field basis with diagonal $Y^u=\hat Y^u$, $Y^d=V\hat Y^d$, 
and 
$\tw=\hat\tw$. Our approximation consists in assuming that all components of the latter are of the same order, and 
again 
taking $c_1=c_2=c_3=1$. Under such assumptions one finds that
\eqali{
\label{thetabar-invs}
I_1&\sim |\tw_3\tw_2^*|\lambda^2_C \frac{g'^2}{16\pi^2}\frac{y^2_t}{16\pi^2}\frac{y^2_b}{16\pi^2} \sim 5\times 
10^{-12}|\tw_3\tw_2^*|
\sim 1.2\times 10^{-15}|\tw|^2
\\
I_{2}&\sim |\tw_3\tw_2^*|\lambda^2_C  \left(\frac{y^2_t}{16\pi^2}\right)^2\frac{y^2_b}{16\pi^2} \sim 3\times 
10^{-13}|\tw_3\tw_2^*|
\sim 7.5\times 10^{-19}|\tw|^2
\\
I_{3}&\sim |\tw_3\tw_2^*||\tw_3|^2\lambda^2_C  \left(\frac{y^2_t}{16\pi^2}\right)^2\frac{y^2_b}{16\pi^2} \sim 3\times 
10^{-13}|\tw_3\tw_2^*||\tw_3|^2
\sim 1.8\times 10^{-21}|\tw|^4,
}
where $\lambda_C\sim0.2$ is the Cabibbo angle and in the final step we used the rough approximation in 
\eqref{wtilde:hierarchy}. We have retained the parametric dependence on the norm of $\tw$, since such a vector is 
actually unbounded and has usually a norm much larger than unity (see Section \ref{sec:hierarchies}). Of course, the 
requirement that the coupling $Y^{Qu}$ stays perturbative also constrains the size of $|\tw|^2$, see \eqref{WR:MB}. Our 
estimates suggest that $I_3$ should be more relevant than $I_2$, and potentially not far from the current experimental 
bound. This qualitative behavior is confirmed by the numerical analysis presented below.

%%%%%%%%%%%%%%%%%%%%%%%%%%%%%%%%%%%%%%%%%%%
\subsection{Numerical estimates}

Here we estimate numerically the irreducible contributions due to the three dominant flavor 
invariants $I_1,I_2$ and $I_3$ in \eqref{Inv1}, \eqref{Inv2} and \eqref{Inv3}. 

The coefficients $c_1,c_2,c_3$ in front of the invariants are expected to be of order one and can be only determined by 
a full calculation within a complete model. In the plots we take $c_1=c_2=c_3=1$ for simplicity.

\begin{figure}[h!]
\includegraphics[scale=0.75]{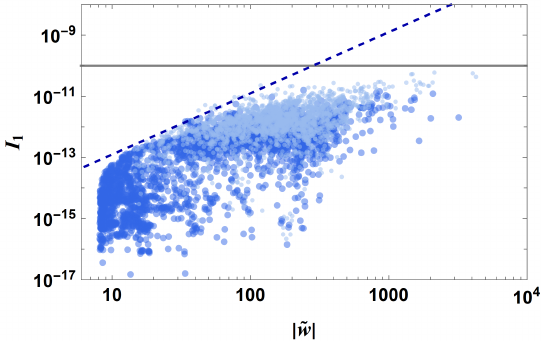}
\includegraphics[scale=0.75]{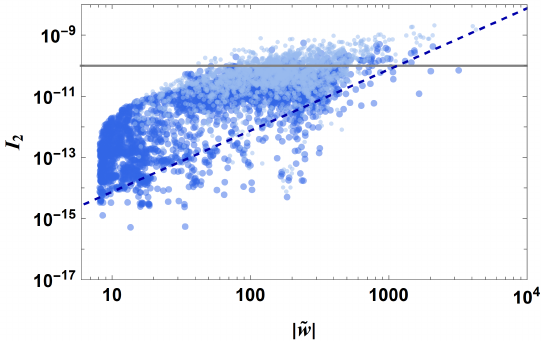}
\caption{\label{fig:I1.I2}%
Contributions to $\btheta$ as a function of $|\tw|$ parametrized by the invariants $I_1$ and $I_2$ in  
\eqref{Inv1} and \eqref{Inv2} for $c_1=c_2=1$. The lighter blue points are excluded by the oblique 
parameter constraints for $M_Q = 8\,\text{TeV}$ while the blue points are allowed by all other constraints.
The dashed blue line represents the estimate in \eqref{thetabar-invs}, and the solid gray line 
denotes the experimental bound \eqref{thetabound}. All points satisfy the perturbativity bound \eqref{cYu<3}.
}
\end{figure}

\begin{figure}[h!]
\includegraphics[scale=0.75]{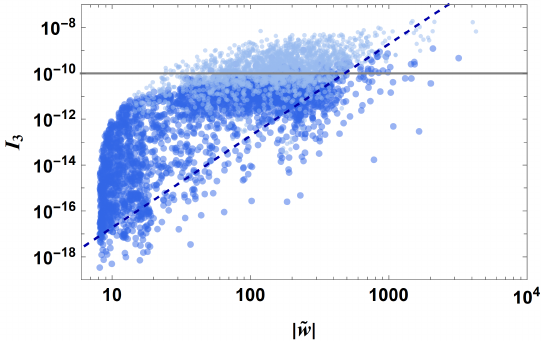}
\caption{\label{fig:I3}%
The contribution to $\btheta$ as a function of $|\tw|$ from the 
invariant $I_3$ in \eqref{Inv1} for $c_3=1$.
The rest is as in Fig.\,\ref{fig:I1.I2}.
}
\end{figure}

In Fig.\,\ref{fig:I1.I2} we show $I_1$ and $I_2$, while in Fig.\,\ref{fig:I3} we show $I_3$. The experimental bound  
\eqali{
\label{thetabound}|\bar\theta|\lesssim10^{-10}.
}
is included as a horizontal grey line. We can see that $I_3$ likely corresponds to the largest contribution to 
$\btheta$. In the plots we only show points that satisfy the perturbativity bound \eqref{cYu<3}.

It is interesting to compare the impact of \eqref{thetabound} on the parameter space compared to that arising from the 
electroweak constraints. This is seen in Figs. \ref{fig:I1.I2} 
and \ref{fig:I3}, where the lighter color denotes the points that are already excluded by the $T$ parameter with 
$M_Q=8\,\unit{TeV}$. We see that the latter constraint is comparable to \eqref{thetabound} in the case of TeV scale 
VLQs, though of course they should 
be viewed as complementary. Crucially, however, for $M_Q>8$ TeV the oblique parameter constraints become weaker whereas 
the one from \eqref{thetabound} does not decouple. For large VLQ masses, that is clearly the most important constraint 
on our models.

%%%%%%%%%%%%%%%%%%%%%%%%%%%%%%%%%%%%%%%%%%%
\section{Conclusions}
\label{sec:conclusions}

In this work we performed a systematic study of Nelson--Barr (NB) solutions to the strong CP problem featuring 
vector-like partners of the SM quark doublet. These models have received considerably less attention than their 
analogues with electroweak-singlet VLQs, yet our analysis shows that this neglect is unwarranted.

We first identified the region of parameter space in which the SM flavor structure --- the quark Yukawa couplings and 
the CKM matrix --- is successfully reproduced. Because no small expansion parameter controls this identification, a 
numerical analysis was necessary. 

We then estimated the dominant contributions to $\bar\theta$ arising from \textit{irreducible} diagrams involving both 
the VLQ and the SM fields. A key observation is that these contributions first appear only at three loops, owing to an 
accidental symmetry of the renormalizable model. As a consequence, in a significant portion of the parameter space the 
resulting $|\bar\theta|$ lies well below current bounds yet within reach of future experiments, in particular proposed 
proton EDM measurements~\cite{pEDM:2022ytu}.

We also examined other relevant experimental constraints on these scenarios. The exotic couplings $Y^{Qu,Qd}_i$ between 
the VLQ and the SM exhibit, in both the up and down sectors, a hierarchical pattern $|Y_1| \ll |Y_2| \ll |Y_3|$, 
inherited from --- though milder than --- the quark-mass hierarchy. This feature distinguishes NB doublet VLQs from 
generic doublet VLQs and ensures that flavor-violating effects beyond the SM are typically small. Overall, within the 
perturbative domain, electroweak precision data provide the leading bounds for VLQ masses below a few~TeV, while 
hadronic CP violation induced by $\bar\theta$ becomes increasingly important at higher masses. Nevertheless, a broad 
region of parameter space remains currently viable, offering meaningful opportunities for future exploration.

In conclusion, our results show that doublet NB--VLQs --- complementing the better-known singlet NB--VLQs of up or down 
type --- constitute a robust and well-motivated solution to the strong CP problem, with potentially interesting 
phenomenological signatures.

%%%%%%%%%%%%%%%%%%%%%%%%%%%%%%%%%%%%%%%%%%%
\appendix
%%%%%%%%%%%%%%%%%%%%%%%%%%%%%%%%%%%%%%%%%%%
\section{Change of basis}
\label{ap:change}

The change of basis from \eqref{lag:Q:NB} to \eqref{lag:Q:gen} can be performed by a unitary transformation on the space 
of doublets $(q_{L},Q_L)$,
\eq{
\mtrx{q_L\cr Q_L}\to W_L\mtrx{q_L\cr Q_L}\,,
}
where $W_L\in U(3+n)$. Here we leave the number $n$ of VLQ doublets general.

The basis change can be performed considering the mass matrices after electroweak symmetry breaking, which read
\begin{equation}\label{MMdown}
    \text{NB}:  \mathcal{M}_d=  \begin{pmatrix}
    \frac{v}{\sqrt{2}}\mathcal{Y}^d&\cM^{qQ}\\
    0&\mathcal{M}_Q
    \end{pmatrix} ;\quad \text{generic}:  M^{q+Q}_d=  \begin{pmatrix}
    \frac{v}{\sqrt{2}}Y^d& 0\\
    \frac{v}{\sqrt{2}}Y^{dQ}&M_Q
    \end{pmatrix},
\end{equation}

\begin{equation}\label{MMup}
    \text{NB}:  \mathcal{M}_u=  \begin{pmatrix}
    \frac{v}{\sqrt{2}}\mathcal{Y}^u&\cM^{qQ}\\
    0&\mathcal{M}_Q
    \end{pmatrix} ;\quad \text{generic}:  M^{q+Q}_u=  \begin{pmatrix}
    \frac{v}{\sqrt{2}}Y^u& 0 \\
    \frac{v}{\sqrt{2}}Y^{uQ}&M_Q
    \end{pmatrix},
\end{equation}
The unitary transformation $W_L$ connects the mass matrices in the two basis:
\eq{
\label{basis:real.gen}
M^{q+Q}_{d/u}=W_L^\dag\cM^{q+Q}_{d/u}\,.
}
We can write an explicit expression for $W_L$ as
\eqali{\label{here}
W_L&=\mtrx{\id_3 & \tw \cr -\tw^\dag & \id_n}
\mtrx{(\id_3+\tw\tw^\dag)^{-1/2} & 0 \cr 0 & (\id_n+\tw^\dag \tw)^{-1/2}}
\,,\\
&=
\mtrx{(\id_3+\tw\tw^\dag)^{-1/2} & \tw (\id_n+\tw^\dag\tw)^{-1/2} \cr -\tw^\dagger (\id_3+\tw \tw^\dag)^{-1/2}& 
(\id_n+\tw^\dag \tw)^{-1/2}}
}
where 
\eqali{
\label{tw:cM}
\tw&=\cM^{qQ}\cM^{Q^{-1}}.
}
Note that both matrices in \eqref{here} are necessary to have a unitary $W_L$, with the first ensuring the zero in the 
upper-right block of $M^{q+Q
}$.

It will be useful for us to re-express all these relations in terms of a different matrix $w$, defined by
\eq{
\label{tw:w}
w=\tw(\id_n+\tw^\dag\tw)^{-1/2}
=(\id_3+\tw\tw^\dag)^{-1/2}\tw\,.
}
In this notation the expressions simplify. In particular:
\eq{
\label{WR:w}
W_L=\mtrx{\big(\id_3-ww^\dag\big)^{1/2} & w\cr -w^\dag & \big(\id_n-w^\dag w\big)^{1/2}}\,.
}
and
\subeqali{
Y^u&= (\id_3+\tw\tw^\dag)^{-1/2}\cY^u=(\id_3-ww^\dag)^{+1/2}\cY^u\,,
\\
Y^d&= (\id_3+\tw\tw^\dag)^{-1/2}\cY^d=(\id_3-ww^\dag)^{+1/2}\cY^d\,,
\\\label{YQuYu}
Y^{Qu}&= \tw^\dagger(\id_3+\tw \tw^\dag)^{-1/2}\cY^u =w^\dag\cY^u={\tw}^\dag Y^u\,,
\\\label{YQdYd}
Y^{Qd}&= \tw^\dagger(\id_3+\tw \tw^\dag)^{-1/2}\cY^d =w^\dag\cY^d={\tw}^\dag Y^d\,,
\\
M^Q&=(\id_n+\tw^\dag \tw)^{+1/2}\cM^Q=(\id_n-w^\dag w)^{-1/2}\cM^Q\,.
}

In the case $n=1$ discussed in the main text, $\tw$ is a vector with norm $|\tw|\geq0$. On the other hand, 
from \eqref{tw:w} it follows that 
\eq{
%\label{|tw|}
|w|^2=\frac{|\tw|^2}{1+|\tw|^2}\,,
}
which says that $0\leq|w|<1$.
CP violation in the SM further excludes $|w|=|\tw|=0$.

%%%%%%%%%%%%%%%%%%%%%%%%%%%%%%%%%%%%%%%%%%%
\section{How to incorporate the SM Yukawas}
\label{ap:inversion}

Here we detail how to solve \eqref{HR} in terms of the eigenvalues, i.e., the SM Yukawas.
After splitting both $\Omega$ and $YY^\dag$ into their real and imaginary parts, one obtains 
\eqref{inversion:re.im}.
The real part can be solved by \eqref{ycal=X}.
Plugging it into the imaginary part yields
\eq{
\label{H2-tilde:O}
\Omega_1^{-1/2}\Omega_2\Omega_1^{-1/2}=\cO^\tp Z\cO\,,
}
which specifies the orthogonal matrix $\cO$.

Without loss of generality we can choose the basis \eqref{w:ab} for $w$.
In such a basis
\eq{
\Omega_1+i\Omega_2=\diag(1-a^2,1-b^2,1)+iab\mtrx{&1&\cr -1&&\cr &&0}\,,
}
and the left hand side of \eqref{H2-tilde:O} has the standard form
\eq{
\Omega_1^{-1/2}\Omega_2\Omega_1^{-1/2}=
\mu\mtrx{&1&\cr -1&&\cr &&0}\,,
}
with $\mu$ given by \eqref{mu=a.b}.
The right hand side of \eqref{H2-tilde:O} tells us that $\cO$ is the matrix that transforms $Z$ to the standard form 
above 
and the third column in $\cO$ is the eigenvector associated to zero eigenvalue of $Z$.
The orthogonal matrix $\cO$ is defined up to an additional rotation in the $(12)$ space from the right.

Eq. \eqref{H2-tilde:O} applies to both up and down Yukawas, in general with ${\cal O}_u\neq{\cal O}_d$ and $\bz^u\neq 
\bz^d$. However, since ${\text{Tr}}[Z^2]=-2|\bz|^2$ is determined entirely by $\Omega$, it must be the same in both 
sectors. The following nontrivial relation $\mu_u=\mu_d=\mu$, where $\mu_u=|\bz^u|$ and $\mu_d=|\bz^d|$, then holds.
This is what we anticipated in \eqref{muu=mud}.
From the definition, we must have $0\le \mu\le 1$\,\cite{consequences}.
These upper and lower limiting values are unphysical.
The limit $\mu=0$ corresponds to the CP conserving case where ${Y^u}^\dag Y^u$, ${Y^d}^\dag Y^d$ are both real.
The other limit $\mu=1$ corresponds to massless quarks as \eqref{HR:Z} would have vanishing determinant.
Reproducing the eigenvalues of the SM Yukawas at 1\,\unit{TeV} requires
\eq{
\label{doublet:mu-constraint}
1-\mu_u> 1.1\times 10^{-10}\,,\quad
1-\mu_d> 1.9\times 10^{-6}\,.
}
See appendix \ref{ap:max:mu} for details.

The vector $\bz=(z_i)$ is an eigenvector of $Z$ associated to zero eigenvalue, so the third column of $\cO$
is proportional to it.
The parameter $\mu$ is the norm $|\bz|$.
Restricted to $\det\cO=1$, we can conventionally choose the third column of $\cO$ as $\bz/\mu$ directly.
Note that $Z\to OZO^\tp$ induces $\bz\to O\bz$.

Now, given $\bz$, the three parameters $x_i$ in \eqref{X:xi}
can be determined from the SM Yukawas $y_i$ ($y_{u_i}$ or $y_{d_i}$).
The relation comes from the characteristic equation for $H_R=Y^\dag Y$,
\eq{
\det(H_R-\lambda I)=-\big[\lambda^3-\gamma_1(H_R)\lambda^2-\gamma_2(H_R)\lambda-\gamma_3(H_R)\big]\,,
}
where the coefficients are
\eqali{
\label{gammai:A}
\gamma_1(H_R)&=\Tr[H_R]=x_1+x_2+x_3\,,\cr
\gamma_2(H_R)&=\ums{2}\Tr[H_R^2-\gamma_1(H_R)H_R]=-x_1x_2(1-z_3^2)-x_2x_3(1-z_1^2)-x_3x_1(1-z_2^2)\,,\cr
\gamma_3(H_R)&=\ums{3}\Tr[H_R^3-\gamma_1(H_R)H_R^2-\gamma_2(H)H]=\det(H_R)=x_1x_2x_3(1-\mu^2)\,.
}
The last expression in each line is obtained with the last expression in \eqref{HR:Z}
where we have used $\mu=|\bz|$.
On the other hand, each of the $\gamma_k(H_R)$ is fixed from the SM Yukawas $h_i\equiv y_i^2$:
\eqali{
\label{gammai:sm}
\gamma_1(H_R)&=h_1+h_2+h_3\,,\cr
\gamma_2(H_R)&=-(h_1h_2+h_2h_3+h_3h_1)\,,\cr
\gamma_3(H_R)&=h_1h_2h_3\,.
}
Because $\Tr[H_R]=\Tr[\re(H_R)]$ but the determinant obeys $\det[H_R]< \det[\re(H_R)]$, the spectrum of $\re(H_R)$ is 
squashed compared to the spectrum of $H_R$.
This is confirmed in Fig.\,\ref{fig:x_i-mu}. Comparison between $\det H_R$ and $\det\re(H_R)$ also leads to a nice 
formula for $\mu$ depending 
solely on $H_R$\,\cite{nbvlq:more}:
\eq{
\label{mu:det}
\mu=\sqrt{1-\frac{\det Y^\dag Y}{\det\re(Y^\dag Y)}}
\,.
}

We now analyze the possibilities for $z_i$.
We first note that the expressions \eqref{gammai:A} are insensitive to the sign of $z_i$.
In the basis where $X$ is diagonal, we can still keep the form \eqref{HR:Z} for $H_R$ if we apply
sign flips to $d_R$ (or $u_R$) which induces sign flips to both sides of $H_R$.
This leaves $x_i$ unchanged but sign flips are induced on $z_i$. Since $\bz$ is a pseudo-vector due
to \eqref{def:Z} we can at most equate the sign of all components and choose either
$(z_1,z_2,z_3)\sim (+++)$ or $(---)$.
With the convention adopted for $\cO\in SO(3)$, these signs are also the same for the third column of $\cO$.

We can end this part checking the number of parameters.
If we name the columns of $\cO$ explicitly as 
\eq{
\cO=
\left(
\begin{array}{c|c|c}
\br_1 & \br_2 & \br_3 
\end{array}
\right)\,,
}
we have $\br_3=\bz/\mu$.
So $\cO$ has its third column defined by $\bz$ while $\br_1,\br_2$ are orthogonal.
The direction of $\bz$ can be parametrized by two angles while the two directions in the orthogonal 
plane depend on one more angle.
So $\cO$ depends on three angles as usual.
The 15 parameters are listed in \eqref{parameters}.
The directions of the vectors $\bz^u$ and $\bz^d$ are defined by the third column of $\cO_u,\cO_d$ while 
their norm should coincide with $\mu$.
With $\bz^u$ and $\bz^d$, the eigenvalue equations \eqref{gammai:A} determine $x^u_i, x^d_i$ given the
Yukawas $y_{u_i},y_{d_i}$.
This fixes six parameters.
Leaving $\cM^Q$ aside, there are still 8 free parameters in $(\mu,b)$, $\cO_u,\cO_d$.
Four of them should be determined from the CKM structure.

%%%%%%%%%%%%%%%%%%%%%%%%%%%%%%%%%%%%%%%%%%%
\section{Constraint on $\mu$}
\label{ap:max:mu}

\begin{figure}[h]
\includegraphics[scale=0.48]{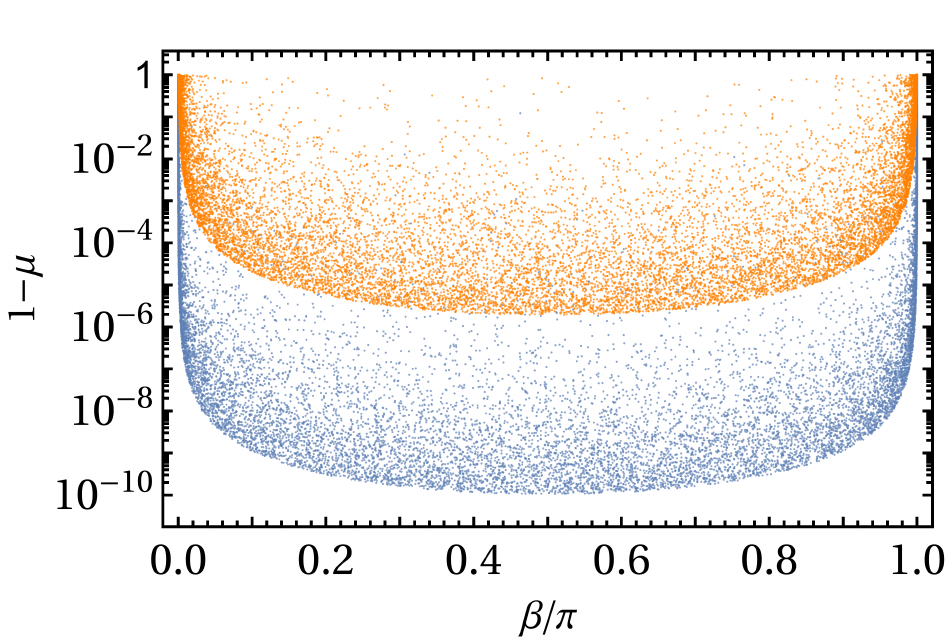}
\caption{\label{fig:mu-beta}%
$1-\mu$ in \eqref{mu:det} as a function of $\beta$ in \eqref{V:1-phase} for up (blue) and down (orange) sectors.
}
\end{figure}

Let us decompose
\eq{
Y^\dag Y= V_R\hY^2 V_R^\dag\,,
}
with $\hY$ being the diagonal matrix,
and consider the one-phase parametrization of a unitary matrix with rephasing freedom from the right\,\cite{one.phase}.
After removing an orthogonal matrix from the right, we can parametrize the remaining freedom as
\eq{
\label{V:1-phase}
V_{R}^\dag=R_{23}R_{13}\diag(1,1,e^{i\beta})\,.
}
We can then calculate the value of $\mu$ using eq.\,\eqref{mu:det}.
Fig.\,\ref{fig:mu-beta} shows the possible values for $1-\mu$ for the up and down sectors using best fit values for the 
SM Yukawas at 1 TeV and arbitrary values for the angles and the phase $\beta$ in \eqref{V:1-phase}.
The non-vanishing of the SM quark masses requires that $\mu_u$ and $\mu_d$ cannot be arbitrarily close to unity.
The figure leads to the constraint \eqref{doublet:mu-constraint} and, to be equal, both $\mu_u,\mu_d$ need to obey the 
more strict bound for $\mu_d$.

Notice that these loose bounds follow solely from the non-vanishing quark masses as $V_R$ for the righthanded fields 
$u_R,d_R$ are not fixed by SM structure.
Imposition of the SM CKM structure including CP violation further restricts $\mu$.
For example, the CP conserving limit $\mu=\mu_u=\mu_d=0$ is not physical.
The physical range restricted additionally by the CKM is shown in the dark points in Fig.\,\ref{fig:x_i-mu} and shown 
in \eqref{range:1-mu}.

For comparison, for one singlet NB-VLQ\,\cite{consequences} of either up-type or down-type, the constraint on $\mu$ also 
comes from the lefthanded field transformation on $q_L$
which is fixed by the CKM and then $\mu$ is bounded from both sides:
\subeqali{
2\times 10^{-7}\lesssim 1-\mu_u &\lesssim 0.0034\,,
\\
0.0045\lesssim 1-\mu_d &\lesssim 0.43\,.
}
We can see that the range in \eqref{range:1-mu} lies in between the above ones.

%%%%%%%%%%%%%%%%%%%%%%%%%%%%%%%%%%%%%%%%%%%
\section{Using invariants}
\label{sec:invs}

The task of enforcing the CKM mixing to the relations \eqref{YYdag:param} is still a difficult task.
After enforcing the six SM Yukawas, four parameters in \eqref{parameters} should be fixed from the CKM while another 
four parameters remain free.
Although feasible, a scan in a 8-dimensional space is not very efficient.
Part of the difficulty in dealing with \eqref{YYdag:param} is that we cannot work in a basis where
one of them is diagonal.
One way to extract only the information that is weak basis independent is to use flavor invariants.
The use of the Jarslkog invariant to quantity CP violation is a well-known example.
Here we propose a different set of invariants that will prove to be useful.

We first define the shorthands
\eq{
H_u\equiv Y^u{Y^u}^\dag\,,\quad
H_d\equiv Y^d{Y^d}^\dag\,,
}
for the SM Yukawas and
\eq{
\cH_u\equiv \cY^u{\cY^u}^\tp\,,\quad
\cH_d\equiv \cY^d{\cY^d}^\tp\,,
}
for real CP conserving Yukawas.
Then \eqref{Yd.Yu} can be compactly written as
\eq{
\Omega^{1/2}\cH_u\Omega^{1/2}=H_u\,,\quad
\Omega^{1/2}\cH_d\Omega^{1/2}=H_d\,,
}
with the same $\Omega$ for both sectors.

It is immediate to see that the following invariants depend on the CKM and are independent of $\Omega$:
\eqali{
I_{ud}&=\aver{H_uH_d^{-1}}&=\aver{\cH_u\cH_d^{-1}}\,,
\cr
I_{du}&=\aver{H_dH_u^{-1}}&=\aver{\cH_d\cH_u^{-1}}\,.
}
Writing $\cY$ in terms of $X$ using the inversion formula \eqref{ycal=X}, we can write
\eqali{
\label{Iud:O}
I_{ud}&
=\sum_{ij}\frac{x^u_i}{x^d_j}O_{ij}^2
=\frac{x^u_3}{x^d_1}O_{31}^2+\cdots
\,,
\cr
I_{du}&
=\sum_{ij}\frac{x^d_i}{x^u_j}O_{ji}^2
=\frac{x^d_3}{x^u_1}O_{13}^2+\cdots
\,,
}
where we have defined the real orthogonal matrix
\eq{
O\equiv \cO_u\cO_d^\tp\,.
}
In the last equalities in \eqref{Iud:O}, we have written the dominant terms assuming $x^u_i,x^d_i$
are hierarchical and $O_{13},O_{31}$ unsuppressed.
Independently of that, given that all the terms in the expression are positive semidefinite, we can write the bounds
\eqali{
|O_{31}|&<\frac{x^d_1}{x^u_3}\sqrt{I_{ud}}\,,\cr
|O_{13}|&<\frac{x^u_1}{x^d_3}\sqrt{I_{du}}\,.
}
Using the $3\sigma$ ranges shown in table \ref{tab:I-range} and taking the ranges \eqref{xi:ranges} for $x^u_i,x^d_i$,
we obtain 
\eq{
|O_{31}|\lesssim 0.2\,.
}
The bound we obtain for $|O_{13}|$ is larger than unity and hence not useful.
\begin{table}[h]
\[
\begin{array}{|c|c|c|c|}
\hline  
\text{Invariant}    & \text{$3\sigma$ range} & \text{Invariant}    & \text{$3\sigma$ range}
\cr
\hline
\tilde{I}_{ud}=\aver{\tH_u\tH_d^{-1}} & [72.03,114.6] &
\tilde{I}_{du}=\aver{\tH_d\tH_u^{-1}} & [5.86,7.09]\times 10^{5}
\cr
\hline
\tilde{I}_{udud} & [5.12,13.1]\times 10^{3} &
\tilde{I}_{dudu} & [3.43,5.03]\times 10^{11}
\cr
\hline
\tilde{J}_{\rm dim}/i & [1.00,1.57]\times 10^{12}
& &
\cr
\hline
\end{array}
\]
\caption{\label{tab:I-range}
Range for the flavor invariants with Yukawas fixed at best-fit and CKM varying within $3\sigma$ at 
1\,TeV.
}
\end{table}

We can equally consider the powers $\aver{(H_uH_d^{-1})^2},\aver{(H_dH_u^{-1})^2}$.
But instead of them, we can consider
\eqali{
I_{udud}&\equiv \ums{2}I_{ud}^2-\ums{2}\aver{(H_uH_d^{-1})^2}\,,
\cr
I_{dudu}&\equiv \ums{2}I_{du}^2-\ums{2}\aver{(H_dH_u^{-1})^2}\,.
}

For a CP violating measure, we can consider instead of the usual Jarslkog,
\eq{
J=\det([H_u,H_d])=\aver{H_u^2H_d^2H_uH_d-H_d^2H_u^2H_dH_u}\,,
}
the dimensionless (when quark masses are considered) version
\eq{
J_{\rm dim}=\aver{H_uH_dH_u^{-1}H_d^{-1}-H_dH_uH_d^{-1}H_u^{-1}}\,.
}
Within the SM, they are simply related by
\eq{
J_{\rm dim}=\frac{J}{y^2_{u_1}y^2_{u_2}y^2_{u_3}y^2_{d_1}y^2_{d_2}y^2_{d_3}}\,.
}

The dimensionless version is more suitable for writing in terms of $\cH_u,\cH_d$ as
\eq{
J_{\rm dim}=\aver{\cH_u\Omega\cH_d\cH_u^{-1}\Omega^{-1}\cH_d^{-1}}-c.c.,
}
where $c.c.$ denotes the complex conjugate.
Note that $\Omega$ only appear in two places.
In terms of $X_u,X_d$, it becomes
\eqali{
J_{\rm dim}&=i\mu
\left\langle
X_u\mtrx{0&1&0\cr-1&0&0\cr 0&0&0}X_dX_u^{-1}
\mtrx{\frac{1}{1-\mu^2}&&\cr &\frac{1}{1-\mu^2} & \cr &&1}X_d^{-1}
\right\rangle
\cr
&\quad -i\frac{\mu}{1-\mu^2}
\left\langle
X_uX_dX_u^{-1}
\mtrx{0&1&0\cr-1&0&0\cr 0&0&0}X_d^{-1}
\right\rangle
-c.c.
}
Here $X_u,X_d$ are not diagonal and absorbed the matrices $\cO$ as
\eq{
X_u=\cO_u^\tp\hX_u\cO_u\,,\quad
X_d=\cO_d^\tp\hX_d\cO_d\,.
}

We partly use these invariants to guide the search for physical points.

%%%%%%%%%%%%%%%%%%%%%%%%%%%%%%%%%%%%%%%%%%%

%%%%%%%%%%%%%%%%%%%%%%%%%%%%%%%%%%%%%%%%%%%%%%%%%
\end{document}